\documentstyle[12pt,epsf]{article}

\def\ffrac#1#2{\textstyle{#1\over#2}\displaystyle}

\begin{document}
\baselineskip=18pt
\pagestyle{empty}
\begin{center}
{\bf \LARGE Lectures on Conformal Invariance and Percolation$^*$\\}
\vspace{8mm}
{\bf \large John Cardy\\}
\vspace{2mm}
{Department of Physics\\
Theoretical Physics\\1 Keble Road\\Oxford OX1 3NP, UK\\
\& All Souls College, Oxford\\}
\end{center}
\vspace{6mm}
\begin{abstract}
These lectures give an introduction to the methods of conformal field theory as
applied to deriving certain results in 
two-dimensional critical percolation: namely 
the probability that there exists at least one cluster connecting two disjoint
segments of the boundary of a simply connected region; and the mean number of such
clusters. No previous familiarity with conformal field theory is assumed, but in
the course of the argument many of its important concepts are introduced in as
simple a manner as possible. A brief account is also given of some recent
alternative approaches to deriving these kinds of result.
\end{abstract}
\vfill
\noindent$^*$Lectures delivered in ``New Trends of Mathematical Physics
and Probability Theory'', Chuo University, Bunkyo-ku, Tokyo, March 5-6, 2001.
\newpage

\pagestyle{plain}
\setcounter{page}{1}
\setcounter{equation}{0}

\section{Introduction.}
The percolation problem 
has for many years been of great interest to theoretical physicists
and mathematicians, in part because it is so simply stated yet so full
of fascinating results. It embodies many of the important concepts of
critical phenomena, yet is purely geometrical in nature. 

Percolation studies the clustering properties of identical
objects which are randomly and uniformly
distributed through space. In lattice \em bond
\em percolation, the links of a regular lattice, of edge length $a$,
are either \em open \em or
\em closed\em. In the simplest version of the model, the open bonds are
independently distributed with a probability $p$ for each to be open
(and $1-p$ to be closed.) In \em site \em percolation, the bonds are all
assumed to be open, but now each site is open with probability $p$. In
both cases we study the statistical properties of \em clusters \em of
neighbouring open bonds and sites. When $p$ is small, the mean cluster
size is also small, but, in more than one dimension, there is a critical
value $p_c$ of $p$, called the percolation threshold, at which the mean
cluster size diverges. For $p>p_c$ there is a finite probability that a
given site belongs to an infinitely 
large cluster.\footnote{
There are also so-called continuum versions of percolation, for example
the clusters formed by
hard spheres of a given radius $a$ which are distributed independently
so that they may overlap. In all cases, however, 
a finite microscopic length $a$ is necessary to define
the notion of clustering.}

\begin{figure}
\centerline{
\epsfxsize=3in
\epsfbox{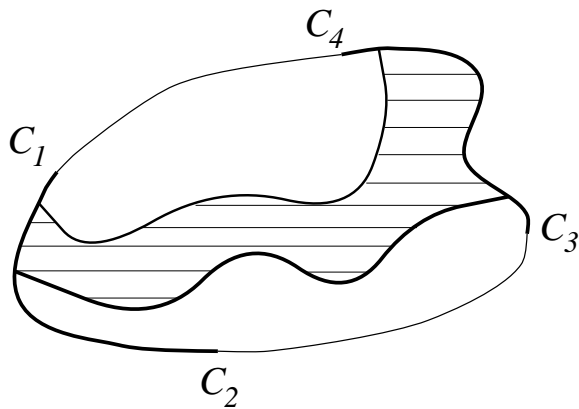}}
\caption{A crossing cluster from $\gamma_1$ $(C_1C_2)$ to $\gamma_2$
$(C_3C_4)$.}
\label{fig1}
\end{figure}

In these lectures, we shall be concerned with properties of the 
\em continuum limit \em of percolation. This may be defined as follows
(we consider two dimensions from now on):
consider a finite region $\cal R$ of the plane, bounded by a curve 
$\Gamma$. Consider a percolation problem on a sequence of lattices ${\cal
L}_a$ covering $\cal R$,
constructed in such a way that the lattice spacing $a\to0$
keeping the size of $\cal R$ fixed. Obviously when we do this, such quantities
as the total number of clusters in $\cal R$ will diverge as $a\to0$, so
we need to identify some suitable quantities which might have a finite
limit. An example is afforded by the \em crossing 
probabilities\em. Suppose for simplicity that $\cal R$ is simply
connected, and let $\gamma_1$ and $\gamma_2$ be two disjoint segments of
$\Gamma$ (see Fig.~\ref{fig1}). Then a \em crossing event \em is a
configuration of bonds (or sites) on the lattice ${\cal L}_a$ 
such that there exists
at least one cluster, wholly contained within $\cal R$,
containing both at least one point of $\gamma_1$
and of $\gamma_2$. 
Let $P(\gamma_1,\gamma_2;{\cal L}_a)$ be the
probability of this event. Then the following statements are all
conjectured to be true:

\begin{itemize}
\item $P(\gamma_1,\gamma_2;{\cal L}_a)$ has a finite limit, denoted by 
$P(\gamma_1,\gamma_2)$, as $a\to0$, which (under a broad class of
conditions) is independent of the particular form of $\cal L$, of the
precise way in which $\cal L$ intersects the boundary $\Gamma$, and of
whether the microscopic model is formulated as bond, site, or any other
type of percolation, as long as there are only short-range correlations
in the probability measure.\footnote{Of course, this limit is
interesting only at $p=p_c$. For $p<p_c$, $\lim_{a\to0}P=0$, because all
clusters are finite (in units of $a$), while
for $p>p_c$ the infinite cluster always spans, so the limit is 1.}
\item $P(\gamma_1,\gamma_2)$ is invariant under transformations of $\cal
R$ which are conformal in its interior (but not necessarily on its
boundary $\Gamma$). 
\item The Riemann mapping theorem allows us to conformally map the
interior of $\cal R$ onto the interior of the
unit disc $|z|<1$ of the complex plane.
Suppose that the ends of the segments are thereby mapped into the points
$(z_1,z_2,z_3,z_4)$ (see Fig.~\ref{fig1} for the labelling). Then
$P(\gamma_1,\gamma_2)$ is a function only of their cross-ratio
\begin{equation}
\eta\equiv{(z_1-z_2)(z_3-z_4)\over(z_1-z_3)(z_2-z_4)}\quad,
\end{equation}
and has the explicit form
\begin{equation}
\label{crossing}
P={\Gamma(\ffrac23)\over
\Gamma(\ffrac43)\Gamma(\ffrac13)}\,\eta^{\ffrac13}\,
{}_2F_1(\ffrac13,\ffrac23;\ffrac43;\eta) 
\end{equation}
where ${}_2F_1$ is the hypergeometric function.
\item Moreover, if the random variable $N_c(\gamma_1,\gamma_2;{\cal
L}_a)$ denotes the total number of distinct clusters which cross from
$\gamma_1$ to $\gamma_2$, then the whole probability distribution of 
$N_c$ also has a finite limit as $a\to0$ and is conformally invariant,
depending only on $\eta$. In particular the mean number of such crossing
clusters is
\begin{equation}
\label{meannumber}
E[N_c]=\ffrac12-\ffrac{\sqrt3}{4\pi}
\left[\ln(1-\eta)
+2\sum_{m=1}^\infty{\Gamma(\ffrac13+m)\Gamma(\ffrac23)
\over\Gamma(\ffrac23+m)\Gamma(\ffrac13)}{(1-\eta)^m\over m}\right]
\end{equation}
\end{itemize}

Equations (\ref{crossing},\ref{meannumber}) are just two among many
similar results which may been obtained using methods of conformal
field theory developed by theoretical physicists\cite{JCcrossing,Watts,
JClink}. At first sight, the
appearance of formulas like these may seem quite mysterious to those
used to thinking about percolation as a lattice problem. It is certainly
true that the methods originally used to derive them are not, so far,
mathematically rigorous.\footnote{S.~Smirnov\cite{Smirnov} has recently
given a proof of (\ref{crossing}) using other methods.}  But there is no doubt
that they are correct - (\ref{crossing}) has been numerically tested to
great precision in a number of cases\cite{lang}, and, moreover, once the existence
of a conformally invariant continuum limit is accepted, the formulas
follow from very classical mathematical methods. 
It is the purpose of these lectures to give some idea to a
non-specialist audience of how these kinds of results arise. Inevitably
I shall not be able to cover all the details, but hopefully the lectures
will provide a basis from which to explore the literature 
further\cite{genref}.\footnote{The main text includes only those concepts and
arguments needed to arrive at the final conclusion. Footnotes will
indicate how these ideas fit within the more general framework of
conformal field theory.} 
For reasons of time I will focus on the derivation of the above formulae,
which is mostly my own work,  as well as mention some other recent
alternative derivations. This is not to overlook the work of others
on other important results in percolation,
particularly that of Nienhuis, Duplantier, Saleur
and others on the
Coulomb gas approach, which is better suited to systems without
boundaries.

\section{Percolation, the random cluster model, and the Potts model.}

The emphasis in this course will be on the analogy between
percolation and conventional critical behaviour in spin systems. This
is through a well-known mapping first discovered by 
Fortuin and Kastelyn\cite{Kast}. The Potts model is a generalisation of
the Ising model in which the spins $s(r)$ at each site of a lattice
take the values $(1,2,\ldots,Q)$, where, initially, $Q$ is an integer
larger than 1. The energy is the sum over all nearest neighbour pairs
$(r',r'')$ 
of sites (i.e. a sum over all bonds) of $-J\delta_{s(r'),s(r'')}$.
Thus the partition function is 
\begin{equation}
Z={\rm Tr}\,\exp(\beta J\sum_{r',r''}\delta_{s(r'),s(r'')})
\end{equation}
Apart from an overall unimportant constant this may be rewritten as
\begin{equation}
Z={\rm Tr}\,\prod_{r',r''}\left((1-p)+p\delta_{s(r'),s(r'')}\right)
\end{equation}
where $p=1-e^{-\beta J}$. 

Now imagine expanding out the product. If there are $B$ bonds on the
lattice there will be $2^B$ terms in this expansion. Each term may be
associated with a configuration in which each bond of the lattice is
open (if we choose the term $\propto p$), or closed (if we choose the
term $\propto (1-p)$). Sites connected by open bonds form clusters, and
the Kronecker deltas force the all the spins in a given cluster to be in
the same state. When we trace over the spins, each cluster will have
only one free spin, so will give a factor $Q$. Thus we can write $Z$
as a sum over configurations $\cal C$ of open bonds:
\begin{equation}
\label{rcm}
Z=\sum_{\cal C}\,p^{|{\cal C}|}(1-p)^{B-|{\cal C}|}\,Q^{N({\cal C})}
\end{equation}
where $N({\cal C})$ is the number of distinct clusters in $\cal C$.
Note that in this form, at least for a finite lattice, 
$Z$ is a polynomial in $Q$ and therefore its definition may be extended
to non-integer values of $Q$.

The weights in (\ref{rcm}) define the \em random cluster model\em.
Of course, in percolation, each cluster is weighted with a factor 1,
so it corresponds to $Q=1$. In that case, the sum is simply over
all possible configurations weighted by their probabilities, 
so $Z(Q=1)=1$. However, there is nontrivial information in the
correlation functions. For example, the probability
that sites $r_1$ and $r_2$ are in the same cluster is given
by the limit as $Q\to1$ of
\begin{equation}
\langle\delta_{s(r_1),a}\delta_{s(r_2),a}\rangle
-\langle\delta_{s(r_1),a}\delta_{s(r_2),b}\rangle
\end{equation}
where $a$ and $b$ are any two different Potts states.\footnote{
This follows from a similar argument to that which gives the crossing
probabilities in terms of partition functions, described later.} 
From (\ref{rcm}) we can also calculate quantities like the mean total
number of clusters by differentiating with respect to $Q$:
\begin{equation}
E[N]=(\partial/\partial Q)|_{Q=1}\,Z(Q)
\end{equation}

It is important establish how quantities like the crossing probabilities
may be related to those in the Potts model. For this we need to bring
in the notion of boundary conditions. Consider a Potts model defined on
a lattice $\cal L$ which covers the region $\cal R$ of the plane, as
described in the Introduction. We consider the boundary of $\cal L$
as being the set of sites which lie just outside $\cal R$ but are
adjacent to sites within $\cal R$. These boundary
sites form a discrete approximation to the boundary $\Gamma$. On these
boundary sites, there are two simple and natural boundary conditions we
might impose on the Potts spins: \em free\em, which means that we sum
freely over them in the partition function; and \em fixed\em, in which
case they are all fixed into some particular Potts state, say $a$.

\begin{figure}
\centerline{
\epsfxsize=4in
\epsfbox{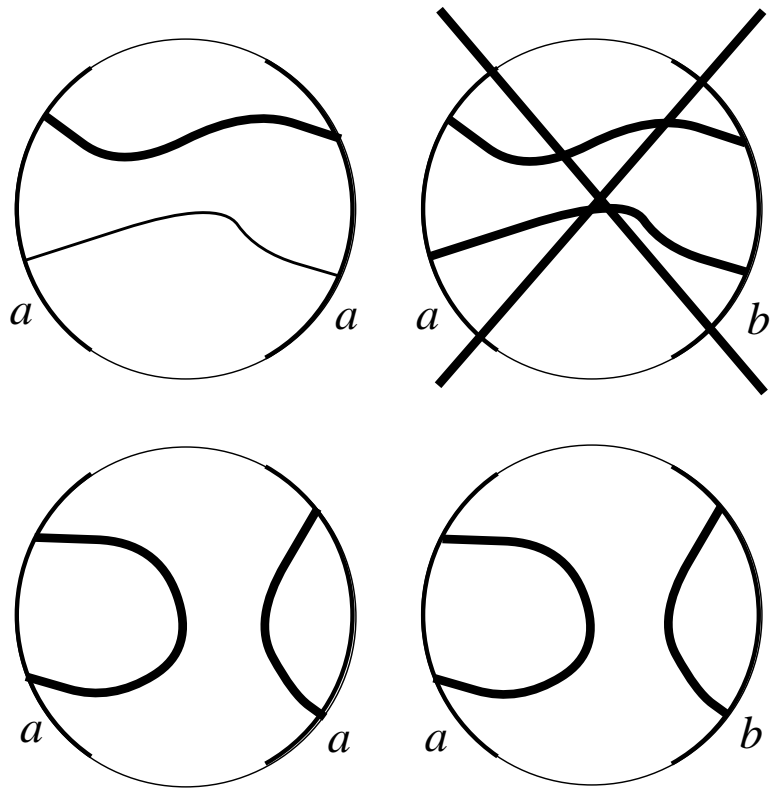}}
\caption{Crossing events contribute to $Z_{aa}$ but not to $Z_{ab}$,
while non-crossing events contribute equally to both.}
\label{fig2}
\end{figure}

Notice that if we take the same boundary condition on the whole
boundary, then in the limit $Q\to1$ partition functions with either free
or fixed boundary conditions become the same,
$Z=1$, since all spins must be in the same state! But we can get
something nontrivial if we allow the boundary conditions to be different
on different parts of the boundary.
In particular, consider the geometry of Fig.~\ref{fig1}, and suppose
the boundary spins are fixed on the segments $\gamma_1$ and $\gamma_2$,
and free on the rest of the boundary. By the permutation symmetry of the
Potts states, there are two possible different cases: when the fixed
states on $\gamma_1$ and $\gamma_2$ are the same, for example $a$; or
when they are different, say $a$ and $b$. Let us denote the partition
functions in the two cases by $Z_{aa}(Q)$ and $Z_{ab}(Q)$ respectively.
Now each configuration $\cal C$ either has, or does not have, a cluster
which spans between $\gamma_1$ and $\gamma_2$ (see Fig.~\ref{fig2}).
Those which do not have a spanning cluster
contribute to both partition functions (with
the appropriate weights), but configurations which do
cannot contribute to $Z_{ab}(Q)$, since the existence of the cluster
would force spins on the two segments to be in the same state, which is
not true by assumption. Thus we have a simple relation for the
crossing probability, on the lattice:\footnote{Note that the crossing probability
in the random cluster model with $Q\not=1$ is not given by the (normalised)
difference $(Z_{aa}-Z_{ab})/Z_{aa}$, since this counts clusters which touch 
$\gamma_1$ and/or $\gamma_2$ with weight 1, rather than $Q$.}
\begin{equation}
\label{ZZ}
P(\gamma_1,\gamma_2;{\cal L})=
\lim_{Q\to1}\left(Z_{aa}(Q)-Z_{ab}(Q)\right)
\end{equation}
Remember that each of these is a polynomial in $Q$, so that 
$Z_{ab}$ makes perfect sense at $Q=1$, even though it might seem that we
need at least two states to define it.\footnote{We can think
of the crossing probability as $P(N_c\geq1)$, where $N_c$ is the number
of crossing clusters. It is also possible to design more complicated
boundary conditions which give $P(N_c\geq n)$ for any integer $n\geq1$.
See \cite{JCcross2}.}

\begin{figure}
\centerline{
\epsfxsize=3in
\epsfbox{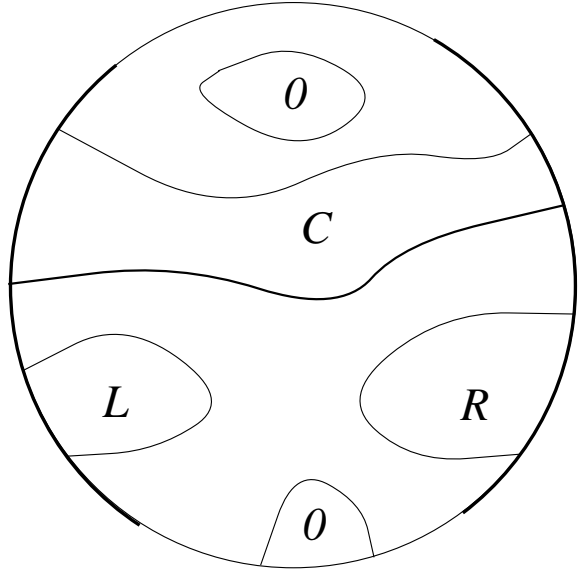}}
\caption{Different types of cluster, which touch both, one or none of
$\gamma_1$ and $\gamma_2$.}
\label{fig3}
\end{figure}

With only slightly more effort we can relate the mean number of distinct
crossing clusters $E[N_c]$ to partition functions. Consider once again
the geometry of Fig.~\ref{fig1}, but now subdivide the clusters into
those which touch neither $\gamma_1$ nor $\gamma_2$, those which touch
$\gamma_1$ but not $\gamma_2$, those which touch $\gamma_2$ but not
$\gamma_1$, and finally the crossing clusters, which touch both 
$\gamma_1$ and $\gamma_2$. Denote the number of each such cluster in each
configuration by $N_0$, $N_L$, $N_R$ and $N_c$ respectively (see
Fig.~\ref{fig3}.) 
Since
clusters which touch a boundary where the Potts spins are fixed are
counted with weight 1, while those which are free are counted with
weight $Q$, it follows that 
\begin{eqnarray}
Z_{ff}&=&\langle Q^{N_c+N_L+N_R+N_0}\rangle\qquad
Z_{aa}=\langle Q^{N_0}\rangle\\
Z_{af}&=&\langle Q^{N_R+N_0}\rangle\qquad\qquad
Z_{fa}=\langle Q^{N_L+N_0}\rangle
\end{eqnarray}
where $f$ now denotes that the boundary spins are free on that portion
of the boundary (they are always free on the complement of
$\gamma_1\cup\gamma_2$.)
Then straightforward algebra shows that
\begin{equation}
\label{Ncav}
\langle N_c\rangle=(\partial/\partial Q)|_{Q=1}
(Z_{ff}+Z_{aa}-Z_{fa}-Z_{af})=
(\partial/\partial Q)|_{Q=1}
(Z_{ff}Z_{aa}/Z_{fa}Z_{af})
\end{equation}
where the last equality holds because all the partition functions
equal 1 at $Q=1$.

It will be the aim in the rest of these lectures to derive explicit
expressions for these partition functions, in the continuum limit,
and for $Q$ close to 1.

\section{The continuum limit of critical lattice models.}

The Potts model in two dimensions is known to have a critical point,
with a divergent correlation length, for $0\leq Q\leq4$.
We now move onto less solid ground. What I am now going to assert is based
on evidence from the exactly solved Ising model $Q=2$, as
well as the analysis of renormalised
perturbation theory near the upper critical dimension.\cite{JCbook}

At the critical point, these theories are believed to be \em scale
invariant\em. What this means for correlation functions is the following.
Consider for example, the correlations 
$\langle s(r_1)s(r_2)\ldots s(r_n)\rangle$
of the local lattice magnetisation in the Ising model. 
Since $s(r)=\pm1$, this quantity has no dimensions.
At the critical point, in the limit where the lattice spacing $a\to0$
with the points $r_j$ kept fixed, it behaves like
$a^{nx}$ times a function of $r_1,\ldots,r_n$, 
where $x$ is a pure number (equal to $\frac18$ for the
Ising model magnetisation.) This means that we can define a 
\em scaling operator \em $\phi(r)\equiv a^{-x}s(r)$ such that the limit
$a\to0$ of the correlation functions 
$\langle\phi(r_1)\phi(r_2)\ldots\phi(r_n)\rangle$ exists. 
The pure number $x$, which we should denote as $x_\phi$, is called the
\em scaling dimension \em of $\phi$. 

Moreover, as long as the original lattice model has sufficient
symmetry under finite rotations, the limit is invariant under
infinitesimal rotations. Thus, far away from any boundaries, the
two-point correlation function $\langle\phi(r_1)\phi(r_2)\rangle$
depends on the separation $|r_1-r_2|$ only, and, on dimensional grounds,
must therefore have the form\footnote{In the case of the magnetisation,
$2x_\phi$ is conventionally denoted
by $(d-2+\eta)$.} 
\begin{equation}
\langle\phi(r_1)\phi(r_2)\rangle={\rm const.}\,|r_1-r_2|^{-2x_\phi}
\end{equation}

The above statement is strictly true only when the points $r_j$ are
distinct, in the limit $a\to0$. That is, they are separated by an
infinite number of lattice spacings. When their relative
distances are kept fixed in units of $a$, other scaling operators may
arise. For example, the product $s(r_1')s(r_1'')$ on neighbouring sites
may be thought of as being proportional to the local energy density
in the Ising model. Correlation functions of this quantity 
behave in a similar way to those of the magnetisation, but they define
a different scaling operator, with a different scaling dimension.
In general there is an infinite number of scaling operators $\phi_k(r)$,
each with their own scaling dimensions $x_k$. Arbitrary local products
$S_j$ of spins which are separated by distances $O(a)$, are given
asymptotically as linear combinations of these scaling operators:
\begin{equation}
\label{S}
S_j(r)=\sum_k A_{jk}a^{x_k}\phi_k(r)
\end{equation}
where the dimensionless coefficients $A_{jk}$ are of order unity, and
correlations of the scaling operators all have a limit as
$a\to0$. These
are rotationally and translationally invariant (far from a
boundary), and also \em scale covariant\em: under a scale transformation
$r\to r'\equiv b^{-1}r$
\begin{equation}
\label{scalecov}
\langle\phi_1(r_1')\phi_2(r_2')\ldots\phi_n(r_n')\rangle
=b^{\sum_kx_k}
\langle\phi_1(r_1)\phi_2(r_2)\ldots\phi_n(r_n)\rangle
\end{equation}
In (\ref{S}) the coefficients depend on the particular lattice and so
on, but the scaling dimensions $x_\phi$ are \em universal\em. For
example, in the Potts model they depend only on $Q$ and the
dimensionality of the system, assuming that the interactions are
short-ranged.\footnote{
One of the important properties of the scaling operators is that they
are complete in the following sense. If, in the above correlation
function, we take the separation $|r_1-r_2|$ to be much smaller
than the distances $|r_1-r_k|$ to the other points with $k\geq3$, this
correlation function may be expressed as a sum over correlation functions
with only a single operator at the mid-point, say, of $r_1$ and $r_2$:
$\langle\phi_1(r_1)\phi_2(r_2)\ldots\rangle
=\sum_kc_{12k}|r_1-r_2|^{x_k-x_1-x_2}
\langle\phi_k((r_1+r_2)/2)\ldots\rangle$
where the dependence on $r_1-r_2$ is dictated by scale covariance
and rotational invariance (we have assumed that all the operators are
scalar under rotations, which is a slight over-simplification).
The important thing is that the coefficients $c_{ijk}$ are also
universal, and in particular independent of the other operators hidden
in the $\ldots$. We can thus remove the $\langle\ldots\rangle$. The
result is called the \em operator product expansion (OPE)\em.}

\subsection{Boundary operators.}
The above statements about the existence of the continuum limit remain
valid in the presence of boundaries. Of course the boundary now breaks
global rotational and translational invariance.
Taking for simplicity the boundary
to run along the real axis, so that the system occupies the upper
half-plane $y>0$, then the two-point function of a scaling operator
has the general form\cite{JCsurf}
\begin{equation}
\langle\phi(x_1,y_1)\phi(x_2,y_2)\rangle 
= (y_1y_2)^{-x_\phi}F\left(|x_1-x_2|/y_1,|x_1-x_2|/y_2\right)
\end{equation}
where $F$ is a scaling function. This is valid for all $y_1$ and $y_2$
strictly positive, in the limit $a\to0$. $F$ behaves in such a way as to
recover the bulk result $((x_1-x_2)^2+(y_1-y_2)^2)^{-x_\phi}$
as both $y_1$, $y_2\to\infty$, but it turns out that the limit as
$y_1$ and/or $y_2\to0$ is singular. In that limit we find instead the
behaviour
\begin{equation}
\label{boundary}
\langle\phi(x_1,y_1)\phi(x_2,y_2)\rangle 
\sim\sum_k B_k^2(y_1y_2)^{-x_\phi+\tilde x_k}|x_1-x_2|^{-2\tilde x_k}
\end{equation}
where the $B_k$ are (universal) constants and
the $\tilde x_k$ are a set of \em boundary scaling dimensions\em.
Thus there is a separate set of scaling
operators $\tilde\phi_k$ defined on the boundary $y=0$, which have 
scaling dimensions $\tilde x_k$
which are in general different from those in the 
bulk.\footnote{(\ref{boundary}) is equivalent to the \em bulk-boundary \em 
expansion $\phi(x,y)=\sum_kB_ky^{-x_\phi+\tilde x_k}\tilde\phi_k(x)$.}
It turns out that the set of such operators depends not only on the bulk
universality class, but also on the nature of the boundary condition,
for example whether the lattice spins are free or fixed.

\subsection{Boundary condition changing operators.}
\begin{figure}
\centerline{
\epsfxsize=4in
\epsfbox{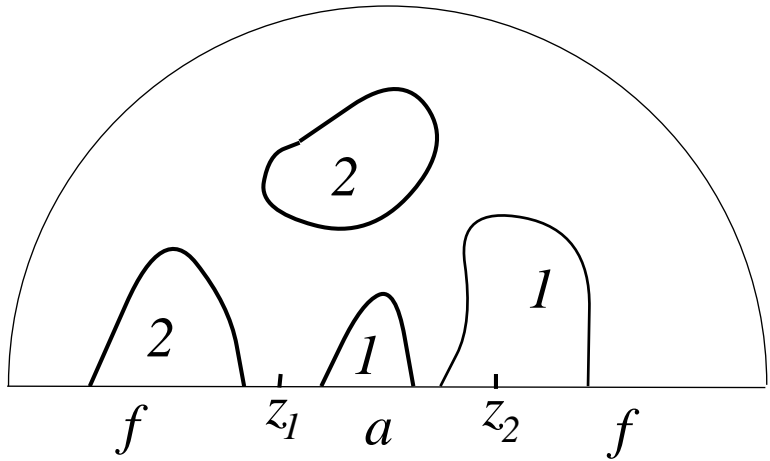}}
\caption{Potts model in the half-plane. The two-point function of the
bcc operator counts clusters which touch the real axis in the interval
$(z_1,z_2)$.}
\label{fig4}
\end{figure}

Once the idea of boundary scaling operators was understood, it was
realised\cite{JCsurf2}
that, at least in two dimensions when the boundary is
one-dimensional, there is another set of objects which should possess
similar properties under scale transformations. These are points on the
boundary where the boundary condition changes from one type to another.
Let us illustrate this with a relevant example from the Potts model.
Consider a Potts model in the half-disc $r<R$, $y\geq0$, see
Fig.~\ref{fig4}. 
Compare the case when (a)
the boundary conditions on the Potts spins are free ($f$) at every point
of the boundary except for the interval $(z_1,z_2)$ of the real axis,
where they are fixed to state $a$, to the case (b) when they are free
everywhere on the boundary. Denote the corresponding partition functions 
by $Z_{af}$ and $Z_{f}$. Then the ratio $Z_{af}/Z_f$ has a finite
limit as $R\to\infty$, \footnote{This is not obvious, at criticality,
and needs to be proved.}
which therefore depends only on $|z_1-z_2|$. We may think of this ratio
as defining the two-point correlation function of 
\em boundary condition changing operators \em (bcc operators) with a
well-defined scaling dimension:
\begin{equation}
\langle\phi_{f|a}(z_1)\phi_{a|f}(z_2)\rangle\equiv
\lim_{a\to0}a^{-2x_\phi}(Z_{af}/Z_f)
\propto |z_1-z_2|^{-2x_\phi}
\end{equation}
Within the Potts model, we expect the scaling dimension of this bcc
operator to be universal, depending only on $Q$. More complicated sets
of boundary conditions may be considered. For example, the partition
function $Z_{ab}$ in (\ref{ZZ}) which we shall need to compute the crossing
probability is related to the correlation function
$\langle\phi_{f|a}\phi_{a|f}\phi_{f|b}\phi_{b|f}\rangle$.

\subsection{Finite-size scaling of the transfer matrix.}
Consider now a lattice model defined on an infinitely long strip
of width $L$,
parametrised by the coordinates $(u,v)$ with $0\leq v\leq L$.
We could consider periodic boundary conditions in
the $v$-direction, but for our purposes it will be more useful to think
about particular boundary conditions, labelled say by $\alpha$ and
$\beta$ (which do not have to be the same) on either edge of the strip.
A convenient way of discussing this geometry is
through the transfer matrix\footnote{I shall attempt to be consistent
and denote true operators by a hat $(\,\hat{}\,)$. Note that the scaling 
`operators'
introduced earlier are not true operators, but
simply local densities, which commute with each other.}
${\hat T}_{\alpha\beta}(L)$, 
which is a finite matrix whose rows and
columns are labelled respectively by the states of the spins in
neighbouring rows $(u,u+a)$, and whose elements are the Boltzmann
weights for the two rows.  If we take the strip to have finite length
$W$ in the $u$-direction, with periodic boundary conditions in that
direction, the partition function is given by
\begin{equation}
Z={\rm Tr}\,{\hat T}^{W/a}=\sum_n\Lambda_n^{W/a}
\end{equation}
where the $\Lambda_n$ are a complete set of eigenvalues of $\hat T$
(which of course depend on $\alpha$, $\beta$ and $L$.)
In this picture the lattice spins $s(u,v)$ themselves
become matrices $\hat s(v)$
(diagonal in this basis),
acting on the same space as does $\hat T$. If the eigenstate
corresponding to the largest eigenvalue of $\hat T$ is denoted by
$|0\rangle$, then a two-point correlation function may be written
\begin{equation}
\langle s(u_2,v_2)s(u_1,v_1)\rangle
=\langle0|\hat s(v_2)
({\hat T}/\Lambda_0)^{(u_2-u_1)/a}\hat s(v_1)|0\rangle
\end{equation}
and so, by inserting a complete set of eigenstates of $\hat T$, we
see that it decays along the strip as a sum of terms of the form
$(\Lambda_n/\Lambda_0)^{(u_2-u_1)/a}$
such that $\langle n|\hat s|0\rangle\not=0$.
Similar equations hold for the correlation function of any local
product $\hat S_j(v)$ of lattice spins.

How do these quantities behave in the continuum limit $a\to0$,
with $L$ fixed? The space of states becomes infinite-dimensional, but
we shall not worry too much about its precise structure. 
The scaling `operators' $\phi(u,v)$ become true local operators 
$\hat\phi(v)$ acting on this space.
It is useful to write the transfer
matrix itself as
\begin{equation}
\hat T_{\alpha\beta}(L)=\exp(-a\hat H_{\alpha\beta}(L))
\end{equation}
where $\hat H$ has eigenvalues $E_n$, so that correlation functions
decay as a sum of terms $e^{-(E_n-E_0)(u_2-u_1)}$. The existence of the
continuum limit implies that the gaps $E_n-E_0$ have a limit as $a\to0$.
Scaling then implies that, at the critical point, each of these gaps
must be proportional to $L^{-1}$. As we shall soon
see, conformal invariance relates the constants of proportionality to
scaling dimensions of operators.

\section{From scale invariance to conformal invariance.}
Let us recall the equation (\ref{scalecov}) for the behaviour of a
general correlation function 
under a scale transformation
$r\to r'\equiv b^{-1}r$:
\begin{equation}
\label{scalecov2}
\langle\phi_1(r_1')\phi_2(r_2')\ldots\rangle
=b^{\sum_kx_k}
\langle\phi_1(r_1)\phi_2(r_2)\ldots\rangle
\end{equation}
One way to understand this equation is from the existence of the continuum
limit $a\to0$, and the absence of any other length scale at the critical
point: if this limit does exist, it should not matter whether we
rescale the lattice spacing in the underlying lattice model by a factor
$b$:
$a\to ba$, in the sense that the measure of those degrees of freedom
which survive the continuum limit should be unchanged. But we can think
of this rescaling (which is a kind of renormalisation group
transformation) as keeping the lattice spacing fixed, and rescaling
$r\to r'\equiv b^{-1}r$.

Now we make the following bold generalisation: since the measure is
ultimately a product over local Boltzmann weights, we may equally well
make a rescaling $a\to b(r)a$ where the factor $b$ now varies
smoothly with $r$ (on a scale $\gg a$.) So we generalise 
(\ref{scalecov2}) to 
\begin{equation}
\label{confcov}
\langle\phi_1(r_1')\phi_2(r_2')\ldots\rangle
=\prod_kb(r_k)^{x_k}
\langle\phi_1(r_1)\phi_2(r_2)\ldots\rangle
\end{equation}
where $b(r)=|\partial r/\partial r'|$, the jacobian
of the transformation.

\begin{figure}
\centerline{
\epsfxsize=3.5in
\epsfbox{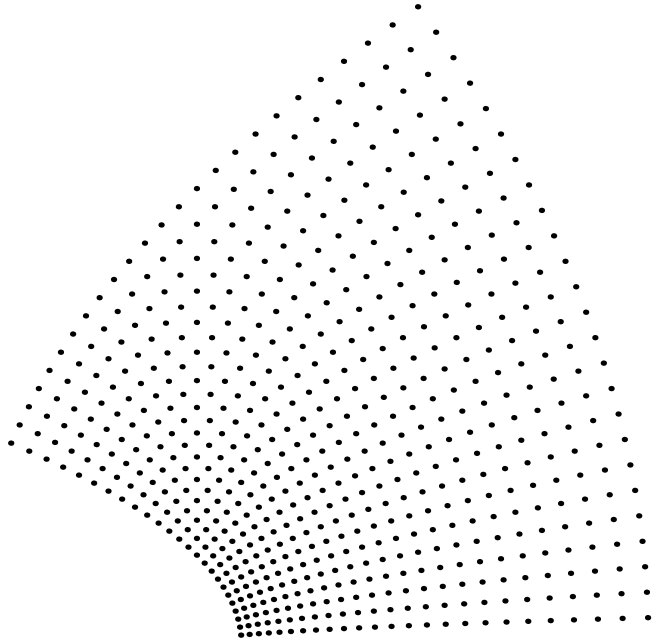}}
\caption{Example of a conformal transformation, locally equivalent to
a scale transformation plus a rotation.}
\label{conffig}
\end{figure}

What are the allowed transformations $r\to r'$? Locally, 
they must correspond to
a mixture of a scale transformation together with a possible rotation
and translation. Such transformations are called \em
conformal\em.\footnote{Strictly speaking, conformal transformations
operate on the metric rather than the coordinates, but this distinction
is not important here.} An example is shown 
in Fig.~\ref{conffig}.\footnote{It is not possible to have 
a non-trivial conformal transformation
without some element of local rotation as well. This means that
anisotropic critical points, which occur, for example, in directed
percolation or at a Lifshitz point, are not conformally invariant.}
So (\ref{confcov}) describes the \em conformal covariance \em of an
arbitrary correlation function.\footnote{This is written for the case of
operators which transform as scalars under rotations. A deeper analysis
also shows that (\ref{confcov}) cannot be true for all scaling
operators: for example it is easy to show that if it is true for $\phi$
it cannot in general be true for $\nabla^2\phi$. However, this analysis
shows that it is true for a subset of operators called \em primary\em,
and the transformation laws for all other operators may be derived from
these.}
Once again, it is not rigorously founded, but rather abstracted from
exactly solved cases like the Ising model, and from the structure of
renormalised perturbation theory. It is generally believed that the
continuum limit of any model which is invariant under scale
transformations, rotations and translations, and has short-ranged
interactions, is also conformally invariant. We are going to assume that
it is valid for the continuum limit of the
critical Potts model with $0\leq Q\leq4$.

In two dimensions, conformal invariance is particularly powerful,
because if we label the points of the plane by a complex number
$z=x+iy$, any analytic function $z\to z'=f(z)$ defines a conformal
transformation, at least at points where $f'(z)\not=0$. 

\subsection{Some simple consequences.} 
Let us consider for definiteness correlations of boundary operators,
since it is these we need for our problem. They may be ordinary limits
of bulk operators at the boundary, or they may be bcc operators.
Just as scale covariance fixes the form of the two-point functions,
conformal invariance gives further information about the higher-point
functions. If we consider the upper half plane, then the set of
conformal transformations which preserve this are the real M\"obius
transformations
\begin{equation}
\label{mob}
z\to z'={az+b\over cz+d},\qquad\quad ad-bc\not=0,
\end{equation}
where $a$, $b$, $c$ and $d$ are real (this is to preserve the real
axis.) Given a 4-point function 
$\langle\phi(z_1)\phi(z_2)\phi(z_3)\phi(z_4)\rangle$, with the $z_j$ real,
it is therefore possible in general to choose the parameters in
(\ref{mob}) such that $z_2$, $z_3$ and $z_4$ are mapped into
three pre-assigned points, say, $(0,\infty,1)$,
corresponding to $z'=(z-z_2)(z_4-z_3)/(z-z_3)(z_4-z_2)$.
However there is always an invariant of
such a transformation: it is the cross-ratio (anharmonic ratio)
\begin{equation}
\eta\equiv (z_{12}z_{34}/z_{13}z_{24})
\end{equation}
(where the notation $z_{kl}\equiv z_k-z_l$ has been introduced)
so that, in this example, $z_1$ gets mapped into $\eta$. 
We may now apply the transformation formula (\ref{confcov}). The
correlation function on the left is some (as yet) unknown function of
$\eta$, and it is simply a matter of working out the jacobian. After
some algebra,\footnote{This is the simpler case when all four operators
have the same scaling dimension. A more general result also holds.}
\begin{equation}
\label{4pt}
\langle\phi(z_1)\phi(z_2)\phi(z_3)\phi(z_4)\rangle
=\left({z_{13}z_{24}\over z_{12}z_{23}z_{34}z_{14}}\right)^{2x_\phi}
\,F(\eta)
\end{equation}

\subsection{Application to crossing probabilities.}
We are already in a position where we can deduce part of the main
result. We begin with the partition functions corresponding to the
geometry of Fig.~\ref{fig1}, in which the Potts spins on the segments
$\gamma_1$ and $\gamma_2$ are fixed, and the rest are free. As we showed
earlier, the crossing probability is given by the limit as $Q\to1$
of the partition functions $Z_{aa}$ and $Z_{ab}$. According to our
analysis above these may be written, as $a\to0$,
\begin{eqnarray}
Z_{aa}/Z_f&\sim& a^{4x(Q)}
\langle\phi_{f|a}(z_1)\phi_{a|f}(z_2)\phi_{f|a}(z_3)\phi_{a|f}(z_4)
\rangle\\
Z_{ab}/Z_f&\sim& a^{4x(Q)}
\langle\phi_{f|a}(z_1)\phi_{a|f}(z_2)\phi_{f|b}(z_3)\phi_{b|f}(z_4)
\rangle
\end{eqnarray}
Here we have introduced
the scaling dimension $x(Q)$ of the $({\rm free}|{\rm fixed})$ bcc operator,
which will depend on $Q$, but not on $a$ or $b$.
Now the Riemann mapping theorem assures us
that there exists a conformal mapping of the interior of the region
$\cal R$ to that of the unit disc, and thence, by a simple M\"obius
transformation, to the upper half plane $y>0$. 
Thus we may use (\ref{confcov}) to relate the ratios of partition
functions in the two geometries.\footnote{In doing this we must assume
that the transformation is conformal also at the points $z_j$. This
requires that the boundary $\Gamma$ be differentiable at that point.
There are interesting additional factors when the boundary has a corner
at one of these points (as happens in the relevant case of the
rectangle). However all these factors are raised to a power proportional
to $x(Q)$, so are not important at $Q=1$. See Ref.~\cite{JCcorners}.}
However, this will bring in non-trivial factors from the jacobian,
evaluated at the points $z_j$, raised to the power $x(Q)$, so that the
ratio of partition functions is \em not\em, in general, conformally
invariant.

However, a remarkable thing happens at $Q=1$. Consider the two-point
function $\langle\phi_{f|a}(z_1)\phi_{a|f}(z_2)\rangle$ in the upper
half plane geometry. As discussed earlier,
this is proportional to the ratio of correlation
functions $Z_{a}/Z_f$, and, at the critical point, decays as
$|z_1-z_2|^{-2x(Q)}$. In the random cluster model, $Z_{a}$ counts
all clusters which touch the real axis in the interval $(z_1,z_2)$ with
a weight 1, while counting all other clusters with weight $Q$ (see
Fig.~\ref{fig4}.) Denoting the total number of each type of cluster by
$N_1$ and $N_2$ respectively, we have 
$Z_{a}=\langle Q^{N_2}\rangle$ and $Z_f=\langle Q^{N_1+N_2}\rangle$.
Trivially, then, both partition functions equal one at $Q=1$, so that
their ratio is independent of $z_1$ and $z_2$. Thus
\begin{equation}
x(1)=0
\end{equation}
an innocent but remarkably interesting equation from the point of view
of conformal field theory.\footnote{Note that if we ask for the mean
number $\langle N_1\rangle$ of clusters touching the interval $(z_1,z_2)$ 
we must first differentiate with respect to $Q$. The result is
$2x'(1)\ln|z_1-z_2|$. As we show later, the prefactor is $\sqrt3/4\pi$.}
From our perspective it means that the crossing probabilities in
percolation (at least those which may be expressed in terms of
$\phi_{f|a}$) are conformally \em invariant\em, in the sense stated in
the Introduction. Thus percolation has stronger properties under
conformal transformations than do more general critical theories. This
may be traced to the fact that its partition function is equal to 1.

\section{Scaling operators and states in conformal field theory.}

We have already discussed the continuum limit of the transfer matrix
on strip of width $L$, and introduced the concept of a state space 
spanned by the eigenvectors of the operator $\hat H$. Now we shall see
that this has a very simple structure due to conformal invariance. 
Consider the upper half plane, with boundary conditions labelled by
$\alpha$ along the real axis. Consider the
conformal mapping
\begin{equation}
z=x+iy\rightarrow w=u+iv=(L/\pi)\ln z
\end{equation}
which maps the upper half plane into such a strip, with boundary
conditions $\alpha$ on both edges. If we apply the transformation law
(\ref{confcov}) to the correlation function
$\langle\tilde\phi(z_1)\tilde\phi(z_2)\rangle$ 
in the half-plane (where $\tilde\phi$
is some boundary operator and we assume that $0<z_1<z_2$), we find 
\begin{equation}
\langle\tilde\phi(u_1,0)\tilde\phi(u_2,0)\rangle
=\left[{\pi/L\over\sinh((\pi/L)|u_1-u_2|) }\right]^{2x_{\tilde\phi}}
\end{equation}
For $|u_1-u_2|\ll L$ this reduces to the same result as in the
half-plane, but in the opposite limit it decays as $\exp(-(\pi\tilde
x_\phi/L)
|u_1-u_2|)$. This means that there must be an eigenstate of $\hat H$
with a gap $E_n-E_0=\pi\tilde x_\phi/L$. In fact it may be argued that to each
boundary scaling operator $\tilde\phi$ there is an eigenstate. This may
be identified by taking $u_1\to-\infty$ (so that $x_1\to0$), in which
case only the state $\langle n|$ with the lowest value of $E_n$ such that
$\langle n|\hat{\tilde\phi}|0\rangle\not=0$ will contribute. A less
obvious statement is that to each eigenstate there is an operator. This
can be seen by choosing the strip to be in some particular eigenstate
$|n\rangle$ at $u=\pm\infty$ and computing correlation functions on the
strip with this state rather than the vacuum $|0\rangle$. On
transforming these back to the half-plane, they turn out to be exactly
as if we had an operator at the origin, with scaling dimension
$(L/\pi)(E_n-E_0)$. Thus we have one of the most important results of
two-dimensional conformal field theory:\footnote{A more field-theoretic
way to state this is that if $\hat H$ generates translations in $u$
along the strip, then $(L/\pi)(\hat H-E_0)$ generates scale
transformations in the half-plane. Eigenstates of the latter
correspond to scaling operators.} 

\begin{itemize}
\item There is a (1-1) correspondence between the (boundary) scaling
operators and the eigenstates of $\hat H$ in the strip. 
\end{itemize}

In what follows, we shall normalise so that $E_0(L)=0$.\footnote{It is
possible to show that $E_0$ is in general independent of $\alpha$ (in
CFTs in which all the scaling dimensions are non-negative), so this is
consistent. However, there is useful information in the $L$-dependence:
if we normalise so that $E_0(\infty)=0$, then $E_0(L)=-\pi c/24L$, where
$c$ is the central charge of the CFT.} This means that a boundary
scaling operator $\tilde\phi$, inserted near the origin, corresponds to
an eigenstate with $E=\pi x_\phi/L$ in the strip. We may extent this
to bcc operators: if we now consider a strip with different boundary
conditions $\alpha$ and $\beta$ on either edge, this corresponds to an
insertion of the bcc operator $\phi_{\alpha|\beta}$, and so the the
lowest eigenvalue of $\hat H_{\alpha\beta}$ is $\pi
x_{\phi_{\alpha|\beta}}/L$. 

\subsection{Descendent operators.}

Let us consider in more detail the structure of the set of scaling
operators. It is useful first to think about 
infinitesimal conformal transformations (which preserve the upper
half-plane). The simplest ones are the translations, $z\to z+a_{-1}$
(where $a_{-1}$ is an infinitesimal parameter), under which a scaling
operator $\tilde\phi(0)$ transforms according to
\begin{equation}
\tilde\phi(0)\to
\tilde\phi(a_{-1})=\tilde\phi(0)+a_{-1}[L_{-1}\tilde\phi](0)
\end{equation}
where $L_{-1}\tilde\phi$ is nothing but $(\partial/\partial
z)\tilde\phi$. There are also the scale transformations 
$z\to z+a_0z$, under which
\begin{equation}
\tilde\phi(0)\to (1+a_0)^{x_{\tilde\phi}}\tilde\phi(0)\sim
\tilde\phi(0)+a_0[L_0\tilde\phi](0)
\end{equation}
where $L_0\tilde\phi=x_{\tilde\phi}\tilde\phi$.
The statement that the correlations of $\tilde\phi$ satisfy 
(\ref{confcov}) is equivalent to assuming
that for a general infinitesimal conformal
transformation $z\to z+\sum_{n=-1}^\infty a_nz^{n+1}$ which is regular at the
origin,
\begin{equation}
\label{primary}
\tilde\phi(0)\to\tilde\phi(0)+a_{-1}[L_{-1}\tilde\phi](0)
+a_0[L_0\tilde\phi](0),
\end{equation}
with no further terms.

However, a more general conformal transformation of this kind is bound
to have singularities elsewhere in the complex plane (including
infinity). In order to understand the implication of this, let us
consider a more general infinitesimal transformation in the form of
a Laurent series, which may be singular at the origin: 
$z\to z+\sum_{n=-\infty}^\infty a_nz^{n+1}$. No matter
how small we choose the parameters $a_n$ for $n<-1$, for sufficiently
small $z$ this  transformation blows up. We deal with this by imagining
cutting out a small half-disc of fixed radius around the origin, and, inside
this, smoothing out the transformation. By definition it cannot be
conformal in this region. The effect of this from the point of view
of the correlation
functions with other operators far from the origin is to generate 
\em new \em operators, called the \em descendents \em $\tilde\phi$.
Thus in general we have
\begin{equation}
\tilde\phi(0)\to\tilde\phi(0)+\sum_{n=-\infty}^\infty a_n
[L_n\tilde\phi](0)
\end{equation}
Notice that by dimensional analysis the descendent operators 
$L_n\tilde\phi$ have
scaling dimensions $x_{\tilde\phi}+(-n)$. 

We can keep doing this, by examining the properties of these descendent
operators under further infinitesimal conformal transformations. We thus
generate an enormous set of descendent operators 
$L_{n_1}L_{n_2}\ldots\tilde\phi$, with scaling dimensions
$x_{\tilde\phi}+\sum_j(-n_j)$. This, in turn, implies that to each
simple eigenstate of the strip transfer matrix $\hat H$ corresonds a
whole tower of other eigenstates, with eigenvalues spaced by integer
multiples of $\pi/L$.\footnote{These descendent operators do not
necessarily satisfy the condition that $[L_n\phi]=0$ for $n\geq1$, that
is they are not primary and they do not satisfy (\ref{confcov}) without
adding some additional terms. In general, however, they are related to
other descendent operators, with lower scaling dimension, are already
constructed. For example, $[L_1L_{-1}\phi]=2[L_0\phi]$. This algebraic
structure is realised by thinking of the $\hat L_n$ as operators which
act on the states of the transfer matrix. They are in fact the Fourier
components $\hat L_n=\int_0^Le^{\pi inv/L}\hat T(v)dv $ of a special
operator called the stress-energy tensor. They satisfy the Virasoro
algebra 
$[\hat L_n,\hat L_m]=(n-m)\hat
L_{n+m}+\frac{c}{12}n(n^2-1)\delta_{n,-m}$, which plays a very important
role in the algebraic development of conformal field theory.
However, we do not really need this formalism for the
purposes of these lectures.}

\section{Null operators and differential equations.}
In general, with each scaling operator $\tilde\phi$ is associated an infinite
number of descendents. However, they do not all have to be independent,
and in many important applications, they are not. In that case, as we
shall show, the correlation functions of $\tilde\phi$ satisfy simple
linear differential equations. 

Let us consider the case of interest, the bcc operator $\phi_{f|a}$,
with scaling dimension $x(Q)$. Consider the partition function $Z_{fa}$
of an annulus $0\leq v\leq L$, $0\leq u<W$, with periodic boundary
conditions in the $u$-direction, free boundary conditions $f$ on $v=0$
and fixed boundary conditions $a$ on $v=L$. As discussed above, we may
write 
\begin{equation}
\label{abc}
Z_{fa}(Q)={\rm Tr}\,\exp(-W\hat H_{fa}(L))
=\sum_j q^{x_j}
\end{equation}
where $q\equiv e^{-\pi W/L}$ and
the sum is over all scaling operators, and their descendents,
which may occur consistent with the given boundary conditions. The lowest is
$x_0=x(Q)$, corresponding to the operator $\phi_{f|a}$ itself.
Let us examine the first few eigenstates above this, bearing in 
mind the fact that, as $Q\to1$, $Z_{fa}(Q)\to1$. The first consequence
is that, as already noted $\lim_{Q\to1}x(Q)=0$. At $Q=1$ this gives a
contribution 1 to $Z$. All other contributions must therefore cancel
among themselves in this limit.
There is a (unique) operator $L_{-1}\phi_{f|a}$, with scaling dimension
$x(Q)+1$, which gives a contribution $q$ to $Z$ at $Q=1$, which must be
cancelled. The only way is by other primary operator(s) $\tilde\psi$
whose scaling dimensions $x_\psi(Q)\to1$ as $Q\to1$. In fact, there are
candidates for such operators: close to the bcc operator $\phi_{f|a}$
the $S_Q$ permutation symmetry of the Potts model is broken down to
$S_{Q-1}$. The lattice operators $\delta_{s(r),b}$ with $b\not=a$ thus
span a $(Q-2)$-dimensional space. We identify $\tilde\psi$ with the
continuum limit of these operators, with degeneracy $(Q-2)$. Thus
\begin{equation}
Z_{f|a}(Q)=q^{x(Q)}+q^{x(Q)+1}+(Q-2)q^{x_\psi(Q)}+\cdots
\end{equation}
So far, so good.\footnote{
Note that we are ignoring the possible contributions of operators with
non-integer scaling dimensions at $Q=1$. These are believed not to arise,
but in any case would not affect the argument.}
However, at the next level we have 
in principle the operators $L_{-2}\phi_{f|a}$, $L_{-1}^2\phi_{f|a}$
(each with weight 1), and $L_{-1}\psi$, with weight $(Q-2)$.  Together
these would give a contribution $q^2$ to $Z_{af}$. There are
two possible resolutions: either there must be yet another new primary operator
$\psi'$ with dimension $x_{\psi'}(Q)\to2$ as $Q\to1$ and degeneracy
$-1+O(Q-1)$; or the two operators $L_{-2}\phi_{f|a}$ and
$L_{-1}^2\phi_{f|a}$ are not independent. In that case the
combination $L_{-2}\phi_{f|a}-\kappa L_{-1}^2\phi_{f|a}$
(for some number $\kappa$)
corresponds to a \em null state\em, which does not contribute to 
the partition function.
We shall \em assume \em that
the latter solution is chosen.\footnote{This is the weakest point in the
whole argument. However, there is further supporting evidence: at $Q=2$
and at $Q=3$ the same property is known to hold\cite{JCeff}. In addition,
the operator product expansion of two such operators is known, by the
so-called \em fusion rules\em\cite{genref}, to generate another operator
degenerate at level 3: this has been identified much earlier\cite{JCeff}
with the magnetisation operator at a free boundary, which, by duality, 
has the same scaling dimension as the bcc operator $\phi_{a|b}$.}

Let us now go back to the half-plane and examine the consequences of
this for the four-point function 
\begin{equation}
G(z_1,z_2,z_3,z_4)\equiv
\langle\phi_{f|a}(z_1)\phi_{a|f}(z_2)\phi_{f|a}(z_3)
\phi_{a|f}(z_4)\rangle,
\end{equation}
where the points $z_j$ lie on the real axis.
Consider the infinitesimal conformal transformation
\begin{equation}
z\to z'=z+a(z)=z+a_{-2}(z-z_1)^{-1}
\end{equation}
At the points $z_j$ $(2\leq j\leq4)$ this is regular, and the
corresponding operators transform simply according to (\ref{primary}). 
At $z_1$ however it is singular, and must be regularised in the manner
described earlier. This then generates the new operator
$a_{-2}[L_{-2}\phi_{f|a}](z_1)$. We may now equate this to
$\kappa[L_{-1}^2\phi_{f|a}](z_1)=
\kappa(\partial/\partial z_1)^2\phi_{f|a}](z_1)$, where $\kappa$ is an 
(as yet) unknown constant. Setting the total variation with respect to
$a_{-2}$ to zero, we therefore find that $G$ satisfies
the differential equation
\begin{equation}
\label{pde}
\left[\kappa{\partial^2\over\partial z_1^2}+
\sum_{j=2}^4\left({1\over z_j-z_1}{\partial\over\partial z_j}
-{x(Q)\over(z_j-z_1)^2}\right)\right]G(z_1,z_2,z_3,z_4)=0
\end{equation}

This is a linear partial differential equation. However, if we
substitute in the form (\ref{4pt}) (which, recall, followed from the
covariance under M\"obius transformations), we find an ordinary
second-order linear differential equation for $F(\eta)$
of a type very familiar to
mathematical physicists. In fact, it is a hypergeometric equation, with
regular singular points at $(0,\infty,1)$.
A lot of tedious algebra can be avoided by identifying the indices of
this equation, that is, the possible power law behaviour of the general
solution near the singular points. For example, if $F(\eta)\sim
\eta^\alpha$ as $\eta\to0$, this implies that 
$G\sim(z_{12}z_{34})^{\alpha-2x(Q)}(z_{13}z_{24})^{-\alpha}$ in the limit
where $z_{12}z_{34}\ll z_{13}z_{24}$. Substituting into (\ref{pde})
and equating the coefficient of $z_{12}^{\alpha-x(Q)-2}$ to zero gives
\begin{equation}
\label{index}
\kappa(\alpha-2x)(\alpha-2x-1)+(\alpha-2x)-x=0
\end{equation}
Now we know that a particular solution is $G$, which in this limit
should factorise:
\begin{equation}
G\sim\langle\phi_{f|a}(z_1)\phi_{a|f}(z_2)\rangle
\langle\phi_{f|a}(z_3)\phi_{a|f}(z_4)\rangle
\sim z_{12}^{-2x(Q)}z_{34}^{-2x(Q)}
\end{equation}
This is because, 
in that limit, clusters which touch both segments $(z_1,z_2)$ and
$(z_3,z_4)$ are very rare.
Therefore the indicial equation (\ref{index}) must have a solution
with $\alpha=0$. This implies that
\begin{equation}
\kappa={3\over2(2x(Q)+1)}
\end{equation}
The other solution is then $\alpha=\frac13+\frac83x(Q)$. 

A similar analysis may be made of the indicial equations at the other
singular points. At $\eta=1$ we also find $(0,\frac13+\frac83x(Q))$, and
at $\infty$, $(-4x(Q),\frac13-\frac43x(Q))$.\footnote{The indices at
$\eta=\infty$ are conventionally defined by $F\sim\eta^{-\alpha}$. There
is a result that the 6 indices of such an equation must sum to 1. This
explains the appearance of the fraction $\frac13$ when $x(Q)=0$.}

The standard form of the hypergeometric equation is
\begin{equation}
\eta(1-\eta)F''+[c-(a+b+1)\eta]F'-abF=0
\end{equation}
for which the indices are $(0,1-c)$, $(a,b)$ and $(0,c-a-b)$ at
$(1,\infty,0)$ respectively. This we have in our case
\begin{equation}
a=-4x(Q),\qquad b=\ffrac13-\ffrac43x(Q),\qquad c=\ffrac23-\ffrac83x(Q)
\end{equation}
This equation has two independent solutions. The one which is regular
at $\eta=0$ is proportional to
\begin{equation}
{}_2F_1(a,b;c;\eta)=1+{ab\over c}{\eta\over1!}+
{a(a+1)b(b+1)\over c(c+1)}{\eta^2\over2!}+\cdots
\end{equation}
which converges for $|\eta|<1$. 
There are in fact many different ways of writing down two independent
solutions: a particularly useful pair are
\begin{eqnarray}
&&{}_2F_1(a,b;a+b+1-c;1-\eta)\nonumber\\
&&={}_2F_1(-4x,\ffrac13-\ffrac43x;
\ffrac23-\ffrac83x;1-\eta)\qquad{\rm and}\\
&&(1-\eta)^{c-a-b}{}_2F_1(c-b,c-a;c-a-b+1;1-\eta)\nonumber\\
&&=
(1-\eta)^{\ffrac13+\ffrac83x}{}_2F_1(\ffrac13-\ffrac43x,
\ffrac23+\ffrac43x;\ffrac43+\ffrac83x;1-\eta)
\end{eqnarray}
Although we have been considering the four-point function of bcc
operators corresponding to $Z_{aa}$, the differential equation
should hold equally well for that corresponding to $Z_{ab}$.
In fact, the two independent solutions above are (apart from constants)
just the solutions
$F_{aa}$ and $F_{ab}$ corresponding to $Z_{aa}$ and $Z_{ab}$
respectively. This is because $\eta\to1$ corresponds to $z_2\to z_3$.
In the second case $(ab)$, this introduces the non-trivial bcc
operator $\phi_{a|b}$ at $z_2$, so cannot contain a term
$\propto(1-\eta)^0$ (incidentally, we see that this operator must have
dimension $\frac13+\frac83x(Q)$); while in the first case, the opposite
is true. The coefficients may then found from the requirement that, as
$\eta\to0$, $F_{aa}\sim F_{ab}\to1$, and the identity
\begin{equation}
{}_2F_1(a,b;c;1)={\Gamma(c)\Gamma(c-a-b)\over\Gamma(c-a)\Gamma(c-b)}
\end{equation}

We therefore have the final results for the partition functions in the
half-plane (normalised so that $Z_f=1$)
\begin{eqnarray}
Z_{aa}&=&\zeta^{2x(Q)}
{\Gamma(\ffrac23+\ffrac43x)\Gamma(\ffrac13-\ffrac43x)\over
\Gamma(\ffrac23-\ffrac83x)\Gamma(\ffrac13+\ffrac83x)}
{}_2F_1(-4x,\ffrac13-\ffrac43x;\ffrac23-\ffrac83x;1-\eta)\\
Z_{ab}&=&\zeta^{2x(Q)}
{\Gamma(1+x)\Gamma(\ffrac23+\ffrac43x)\over
\Gamma(\ffrac43+\ffrac83x)\Gamma(\ffrac13+\ffrac83x)}
(1-\eta)^{\ffrac13+\ffrac83x}\nonumber\\
&&\qquad\times\,{}_2F_1(\ffrac13-\ffrac43x,
\ffrac23+\ffrac43x;\ffrac43+\ffrac83x;1-\eta)
\end{eqnarray}
where $\zeta=(z_{13}z_{24}/ z_{12}z_{23}z_{34}z_{14})$.

Actually, since we are interested in the limit $Q\to1$, that is
$x(Q)\to0$, these simplify greatly to the order 
needed:\footnote{Notice that the ratio of gamma functions in $Z_{aa}$
simplifies
to $\sin\pi(\ffrac23-\ffrac83x)/\sin\pi(\ffrac23+\ffrac43x)$, so is
easy to expand in powers of $x$.}
\begin{eqnarray}
Z_{aa}&=&\left({z_{13}z_{24}\over
z_{12}z_{23}z_{34}z_{14}}\right)^{2x(Q)}
(1+\ffrac{4\pi}{\sqrt3}x)
{}_2F_1(-4x,\ffrac13;\ffrac23;1-\eta)+O(x^2)\\
Z_{ab}&=&
{\Gamma(\ffrac23)\over
\Gamma(\ffrac43)\Gamma(\ffrac13)}
(1-\eta)^{\ffrac13}{}_2F_1(\ffrac13,
\ffrac23;\ffrac43;1-\eta)+O(x)
\end{eqnarray}

From (\ref{ZZ}), evaluated at $Q=1$,
now follows the main result for the crossing probability
in the half-plane 
\begin{equation}
P((z_1,z_2);(z_3,z_4))=1-{\Gamma(\ffrac23)\over
\Gamma(\ffrac43)\Gamma(\ffrac13)}(1-\eta)^{\ffrac13}
{}_2F_1(\ffrac13,\ffrac23;\ffrac43;1-\eta)
\end{equation}
which may also be written
\begin{equation}
P((z_1,z_2);(z_3,z_4))={\Gamma(\ffrac23)\over
\Gamma(\ffrac43)\Gamma(\ffrac13)}\eta^{\ffrac13}
{}_2F_1(\ffrac13,\ffrac23;\ffrac43;\eta)
\end{equation}
Although this follows from standard identities on hypergeometric
functions, there is a more physical argument, based on duality
in bond percolation:
whenever there is a cluster of open bonds connecting $(z_1,z_2)$ to
$(z_3,z_4)$, there cannot be a cluster of open dual bonds connecting
$(z_2,z_3)$ to $(z_4,z_1)$, and vice versa. Therefore there is either
a cluster of open bonds connecting  $(z_1,z_2)$ to
$(z_3,z_4)$, or a cluster of open dual bonds connecting $(z_2,z_3)$ to
$(z_4,z_1)$. Thus 
\begin{equation}
P(z_1,z_2;z_3,z_4)=1-P(z_2,z_3;z_4,z_1),
\end{equation}
or, in terms of the cross-ratio,
\begin{equation}
\label{duality}
P(\eta)=1-P(1-\eta).
\end{equation}

The expression for the mean number of crossing clusters is slightly
more difficult, because we need to differentiate with respect to
$Q$, that is, with respect to $x$. First we note that $Z_{af}$
and $Z_{fa}$ are equal to $z_{12}^{-2x}$ and $z_{34}^{-2x}$
respectively. Thus the ratio of partition functions in (\ref{Ncav})
may written entirely in terms of $\eta$
\begin{equation}
{Z_{aa}Z_{ff}\over Z_{af}Z_{fa}}=
(1-\eta)^{-2x(Q)}(1+\ffrac{4\pi}{\sqrt3}x(Q))
{}_2F_1(-4x(Q),\ffrac13;\ffrac23;1-\eta)
\end{equation}
to the order required. Thus 
\begin{equation}
\label{123}
E[N_c]=x'(1)\left[-2\ln(1-\eta)+\ffrac{4\pi}{\sqrt3}
-4\sum_{n=1}^\infty{\Gamma(\ffrac23)\Gamma(\ffrac13+n)\over
\Gamma(\ffrac13)\Gamma(\ffrac23+n)}{(1-\eta)^n\over n}\right]
\end{equation}
where the series comes from differentiating ${}_2F_1(a,\ldots)$ with
respect to its first argument $a$ at $a=0$. 

Of course this result depends on the (presently) unknown value of
$x'(1)\equiv({\rm d}x/{\rm d}Q)|_{Q=1}$. 
This may be found\footnote{
In fact, the complete
dependence of $x(Q)$ has been conjectured using Coulomb gas
methods.\cite{JClink}}
by noting that, as $\eta\to0$, 
it is extremely unlikely that there is more than one crossing cluster,
so that $E[N_c]\sim P\sim (\Gamma(\ffrac23)/
\Gamma(\ffrac43)\Gamma(\ffrac13))\eta^{\ffrac13}$. 
Applying Stirling's formula to the coefficients of the series in
(\ref{123}) we see that they behave at large $n$ like 
$(\Gamma(\frac23)/\Gamma(\frac13))n^{-\ffrac43}$, which implies that the
singular part of the sum of the series behaves like
$(\Gamma(\frac23)\Gamma(-\frac13)/\Gamma(\frac13))\eta^{\ffrac13}$
as $\eta\to0$. Equating this to the leading term in $P$ then gives
\begin{equation}
x'(1)=-\left(4\Gamma(\ffrac43)\Gamma(-\ffrac13)\right)^{-1}
=-\sin(4\pi/3)/(4\pi)=\sqrt3/(8\pi)
\end{equation}

This result for $x'(1)$ has a direct physical meaning. Consider once
again the annular geometry, formed by sewing together the ends of a
strip of width $L$ and length $W$. 
The mean number of clusters crossing from one edge to the other is once
again given by a formula like (\ref{Ncav}), where now the partition
functions may be represented in the trace form like (\ref{abc}).
In the limit $W\gg L$ the result is particularly simple, since
$Z_{aa}\sim Z_{ff}\to 1$, while $Z_{af}=Z_{fa}\sim \exp(-\pi
x(Q)(W/L))$.
Thus in this limit,
\begin{equation}
E[N_c]\sim 2\pi x'(1)(W/L) = (\sqrt3/4)(W/L)
\end{equation}
It is to be expected that, for large $W$, $E[N_c]$ should be
proportional to $W$, and therefore by scale invariance should go like
$(W/L)$, but the (irrational) universal coefficient is 
quite surprising.

\section{Other approaches.}
\subsection{Crossing probability in the rectangle and modular invariance.}
\begin{figure}
\centerline{
\epsfxsize=3.5in
\epsfbox{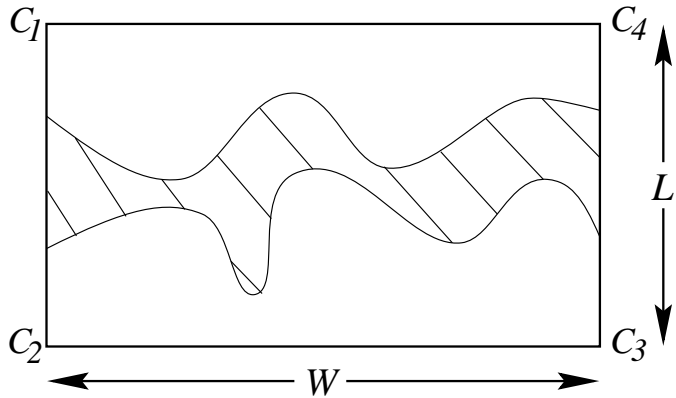}}
\caption{Crossing cluster in the rectangle.}
\label{figrectangle}
\end{figure}

An important special case is when the curve $\Gamma$ is a rectangle,
with the segments $\gamma_1$ and $\gamma_2$ being opposite edges (see
Fig.~\ref{figrectangle}). 
The conformal mapping from the half-plane into
the interior of the rectangle has the form of a Schwartz-Christoffel
transformation
\begin{equation}
z\to w=\int_0^z{dt\over(1-t^2)^{1/2}(1-k^2t^2)^{1/2}}
\end{equation}
which maps the points $(-k^{-1},-1,1,k^{-1})$ to the corners of the
rectangle. The width and height of the rectangle are then
\begin{eqnarray}
W&=&2\int_0^1{dt\over\sqrt{(1-t^2)(1-k^2t^2)}}
=2K(k^2)\\
L&=&\int_1^{1/k}{dt\over\sqrt{(t^2-1)(1-k^2t^2)}}
=K(1-k^2)
\end{eqnarray}
where $K$ is the complete elliptic integral of the first kind.
The \em aspect ratio \em $r\equiv W/L=2K(k^2)/K(1-k^2)$. 
Clearly it is possible numerically to solve this for $k$ for a given
$r$, and compute $\eta=((1-k)/(1+k))^2$ and thence the crossing
probability $P(r)$ via (\ref{crossing}). However, it turns out that there is a
more elegant formula, in terms of modular forms,
due to Kleban\cite{Kleban} (see also
Ziff\cite{Ziff}.) Kleban and Zagier\cite{KZ} have recently derived this
result, using duality and some basic properties of conformal field
theory, with one rather simple assumption, and we now summarise their
argument.

From (\ref{ZZ}), $P(r)=1-Z_{ab}$, where $Z_{ab}$ is the partition
function for a strip of width $L$, with free boundary conditions on each
edge, and length $W$. Think of evaluating this using the continuum
version of the transfer matrix acting horizontally, along the strip:
\begin{equation}
Z_{ab}=\langle a|e^{-W\hat H_{ff}(L)}|b\rangle
\end{equation}
where $\langle a|$ and $|b\rangle$ are so-called \em boundary states\em,
whose form we shall not need. Inserting a complete set of eigenstates
into this matrix element:
\begin{equation}
Z_{ab}=\sum_j\langle a|j\rangle\,q^{x_j/2}\langle j|b\rangle
\end{equation}
where $q=e^{-2\pi(W/L)}=e^{-2\pi r}$ and
the sum is over a complete set of scaling operators, consistent
with the boundary conditions $(ff)$ on the edges, the lowest of which
has $x=0$. The sum may be organised into a sum over primary operators,
and their descendents, so that $P(r)$ has the form
\begin{equation}
\label{pr1}
P(r)=1-\sum_{x \in\pi_{ff}}q^{x/2}Q_x(q)
\end{equation}
where $\pi_{ff}$ is the set of primary scaling dimensions consistent
with the boundary conditions $(ff)$, and $Q_x(q)$ is a power series,
starting with a term $O(q^0)$,
representing the contribution of all the descendents. Note that in this
case no odd powers of $q^{1/2}$ enter. This is a consequence of the
reflection symmetry of the rectangle in the horizontal axis.
Odd powers correspond to states which are odd under this operation, so
do not couple to the states $|a\rangle$ and $|b\rangle$, which are
even. On the other hand, we could also imagine computing $Z_{ab}$
using the transfer matrix acting vertically. The result would
have the form
\begin{equation}
\label{pr2}
Z_{ab}=\sum_{x\in\pi_{ab}}{\tilde q}^{x/2}R_x({\tilde q}^{1/2})
\end{equation}
where $\tilde q\equiv e^{-2\pi/r}$ and 
now odd powers of ${\tilde q}^{1/2}$ are expected to enter because $Z_{ab}$
is not symmetric under reflections in the vertical axis.
Note that the lowest value of $x$ in $\pi_{ab}$ is the scaling
dimension of $\phi_{a|b}$, which is strictly positive.

Finally, the function $P(r)$ also satisfies (\ref{duality}), which means
that $P(r)=1-P(1/r)$. Comparing with (\ref{pr2}) we see that 
$P(1/r)$ has the form of the series on the right hand side, and
comparing this with (\ref{pr1}) we see that in fact we may write
\begin{equation}
P(r)=\sum_{x\in\pi_{ab}}q^{x/2}Q_x(q)
\end{equation}

Now assume that there is only one term in this sum, corresponding
to $x=x_{a|b}$.
Let $r=i\tau$, so that $q=e^{2\pi i\tau}$. Then $P$ is holomorphic in the
upper-half $\tau$-plane. Define $f(\tau)=dP/d\tau$. The modular group is
generated by the elements $S: \tau\to-1/\tau$ and $T:\tau\to\tau+1$,
satisfying $(ST)^3=1$. Now (\ref{duality}) implies that
under $S$, $f(-1/\tau)=-\tau^2f(\tau)$,
while under $T$, 
$f(\tau+1)= e^{i\pi x}f(\tau)$. Since $(ST)^3=1$,
$(e^{i\pi x})^3=-1$, so that $x=m/3$ for some
odd integer $m$. It follows that the function $g\equiv f^6$ satisfies
$g(-1/\tau)=\tau^{12}g(\tau)$, $g(\tau+1)=g(\tau)$, and that it has
a series expansion in positive integers powers of $q$.
It is called a cusp form of weight 12. It turns out
that there is only one such function, up to an overall multiplicative
constant, which is $\eta(q)^{24}$, where $\eta$ is the Dedekind function
\begin{equation}
\eta=q^{1/24}\prod_{n=1}^\infty(1-q^n)
\end{equation}
Thus 
\begin{equation}
P(r)={2^{7/3}\pi^2\over\sqrt3\Gamma(\ffrac13)^3}
\int_r^\infty\eta(ir')^4dr'
\end{equation}
where the constant is fixed by the requirement that $P(0)=1$, and
various identities satisfied by $\eta(q)$. Remarkably, this formula
is identical\cite{Kleban}
to the Schwartz-Christoffel transform of (\ref{crossing}).

\subsection{Carleson's formula for an equilateral triangle, and
Smirnov's proof.}
L.~Carleson made the interesting observation that the crossing formula in an 
equilateral triangle has a much simpler form than that for a rectangle.
This is because of a simple property of the hypergeometric differential equation
arising from
(\ref{pde}): because there is a solution with $F={\rm const.}$, it follows that
$F'(\eta)$ satisfies a simple first order equation which may be solved
by quadrature.
The result is that $P(\eta)$ may be written as
\begin{equation}
\label{int}
P(\eta)={\Gamma(\ffrac23)\over3\Gamma(\ffrac43)\Gamma(\ffrac13)}
\int_0^\eta (t(1-t))^{-2/3}dt
\end{equation}
But the occurrence of the exponent $\frac23$ means that 
this integral has just the form of the Schwartz-Christoffel conformal 
mapping from the upper half-plane to an equilateral triangle!

\begin{figure}
\centerline{
\epsfxsize=3.5in
\epsfbox{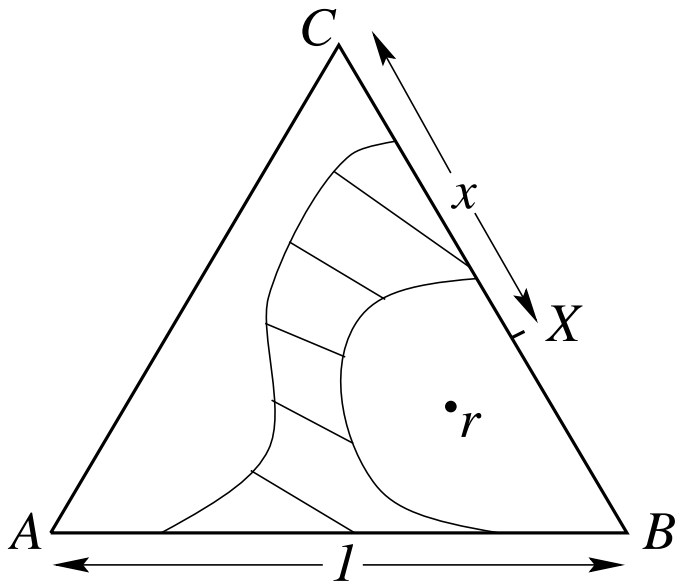}}
\caption{Crossing cluster in the equilateral triangle.}
\label{figtriangle}
\end{figure}

Thus if we consider an equilateral triangle $ABC$ (see Fig.~\ref{figtriangle}),
whose side has a unit length,
and consider the crossing probability from anywhere on side $AB$ to
the segment $XC$ of length $x$ of the side $BC$, then Carleson's version
of the crossing formula is simply 
\begin{equation}
P(AB,XC)=x
\end{equation}

Very recently, S.~Smirnov\cite{Smirnov} has provided a 
\em proof \em of this formula starting from site percolation on the 
triangular lattice. He considers the probability $h(r)$ that a given point
in the interior of the triangle is separated from the segment $AC$ by at
least one cluster spanning from $AB$ to $BC$. Taking $r=X$ on the boundary
segment $BC$ gives the required crossing probability. 
He shows that $h(r)$ is, on the lattice, an approximate harmonic function,
which, in the limit when the lattice spacing $a\to0$, satisfies Laplace's 
equation with the boundary conditions that $h=0$ on $AC$, $h=1$ when
$r=B$ and the components of $\nabla h$ at $60^{\circ}$ to the boundary
(i.e. parallel to $AC$) vanish along $AB$ and $BC$. For the equilateral
triangle, the solution of this is simple: $h(r)$ is the distance of
the point $r$ from the edge $AC$, in units where the height of the triangle
is 1. Of course, in general the solution of this boundary value problem is 
conformally invariant, so in the half-plane, one gets the result
(\ref{int}). Smirnov's arguments explain why the crossing formula
(\ref{crossing}) is the boundary value of an analytic function, but
do not immediately give insight into the conformal field theory
structure underlying this result. It seems to be special to $Q=1$.

\subsection{The approach of Lawler, Schramm and Werner.}
\begin{figure}
\centerline{
\epsfxsize=3.5in
\epsfbox{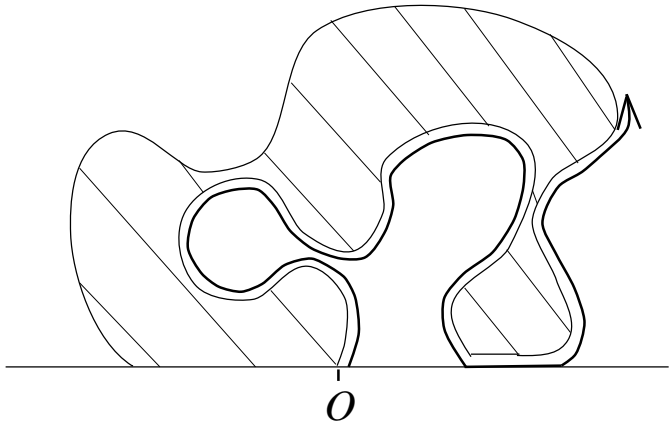}}
\caption{The path which explores the hull of a cluster attached to the
negative real axis. }
\label{figpath}
\end{figure}

Finally we discuss briefly another approach initiated by
Schramm\cite{Schramm}, which fits into a larger investigation 
into the intersection properties of Brownian walks
by the above authors (see
Ref.~\cite{Werner} and many references therein)
The idea is to study the properties
of a percolation hull, or cluster boundary, which grows from the real axis
up into the upper half plane. 
Suppose that all the edges along the negative real
axis are open, and those on the positive real axis are closed. There is an
infinitely large cluster formed by the union of the links on the negative
real axis and those clusters attached to it. Consider exploring the
boundary of this cluster with a time-dependent process which makes one step
along this boundary in unit time. (The process is almost symmetric under
$x\to-x$ because this path is also the boundary of a cluster of open dual
links attached to the positive real axis.) The end of this path executes a
non-crossing random walk in the upper half-plane, occasionally touching 
real axis (see Fig.~\ref{figpath}). 
This successively excludes a larger
and larger region of the half-plane. Schramm argued that the continuum
limit of this process may be defined in terms of the conformal mapping
$g_t(z)$ which maps the complement of the excluded region onto the upper
half-plane. He conjectured\footnote{The recent work of
Smirnov\cite{Smirnov} shows that percolation does indeed have a
conformally invariant continuum limit. Uniqueness
arguments then justify this conjecture \em a posteriori\em.}
that this function satisfies a kind of stochastic
ordinary differential equation
\begin{equation}
\label{stoch}
\partial_tg_t(z)={2\over g_t(z)-a(t)}
\end{equation}
where $g_0(z)=z$ and $a(t)$ is a one-dimensional Brownian motion,
i.e. $\dot a=\zeta(t)$ where $\zeta$ is a gaussian noise satisfying
$\overline{\zeta(t)\zeta(t')}=\kappa\delta(t-t')$. This equation is
defined up to the time that $g_t(z)$ hits $a(t)$: we can think of this
as the time $T_z$ at which the point $z$ enters the excluded 
region.\footnote{It is useful to think about the case
when $a(t)=0$. The solution is then 
$g_t(z)=(z^2+4t)^{1/2}$. The end of the path lies at $2i\sqrt t$
at time $t$. The initial effect of the noise in $a(t)$ is to give this an
additional horizontal motion.} 

The continuum problem is conformally invariant by construction. 
If we now consider the crossing probability between the intervals
$(-\infty,-a)$ and $(0,b)$, it may be related to the times $T_{-a}$
and $T_b$ by
\begin{equation}
P((-\infty,-a),(0,b))=Pr(T_{-a}<T_b)
\end{equation}
From the conjectured equation (\ref{stoch}), with $\kappa=6$,\footnote{Other
values may be considered, which give generalised crossing formulas.
In general, the process generated by (\ref{stoch}) is called Stochastic
Loewner Evolution with parameter $\kappa$ (SLE$\kappa$).}
it may be shown that
the above is given by the crossing formula (\ref{crossing}) with
the cross-ratio $\eta=b/(a+b)$. Of course, many other related formulae
also follow from this approach.

\section{Conclusions.}
The crossing formula (\ref{crossing}) was first conjectured in 1992.
The original argument was based on (1) the mapping to the $Q$-state Potts model
in the limit $Q\to1$ and (2) ideas of conformal field theory which had been
developed up to that time, which have been the main subject of these lectures.
However, it is clear that percolation is somewhat special as in some sense
the measure, suitably defined, is conformally \em invariant\em, rather than
merely covariant as is believed to occur in most critical systems. This
is linked to the fact that for percolation the central charge $c$ 
vanishes.\footnote{In general the partition function of a finite system
in a region size $L$ with smooth metric and smooth boundaries behaves
like $Z\sim e^{c\chi L/6}$ where $\chi$ is the Euler 
character.\cite{JCpesc} Thus the
measure is not strictly even scale invariant when $c\not=0$.}
This invariance has allowed mathematicians to formulate other more direct
approaches to the problem. It remains to be seen whether these methods
will be successful in computing other quantities, such as the mean number
of crossing clusters (\ref{meannumber}), since, at least in our approach, 
this requires going away from $Q=1$, and whether, in a more general
context, they will shed light on the origin of conformal covariance
in more general critical systems.  

{\bf Acknowledgments}. I would particularly like to thank 
Professors Y. Higuchi and M.~Katori for inviting me to Japan to give these
lectures, and for
their kind hospitality. I have benefited from useful correspondence
and conversations with P.~Kleban,  G.~Lawler, O.~Schramm, S.~Smirnov
and W. Werner on this and related subjects, and thank them for sending 
preprints of their work prior to publication. 
This work was supported in part by the EPSRC through grant GR/J 78327.

\end{document}